\documentclass[a4paper,11pt]{article}
\usepackage{pos}

\title{Kaon, Nucleon and $\Delta^*$ Resonances with Hidden Charm}

\author*[a]{Brenda B. Malabarba}
\author[a]{A. Martínez Torres}
\author[b]{K.P. Khemchandani}
\author[c]{Xiu-Lei Ren}
\author[d]{Li-Sheng Geng
}

\affiliation[a]{Universidade de S\~ao Paulo, Instituto de F\'isica,\\ C.P. 05389-970, Sao Paulo, Brazil}

\affiliation[b]{Universidade Federal de S\~{a}o Paulo,\\ C.P. 01302-907, S\~{a}o Paulo, Brazil}
\affiliation[c]{Institut fur Kernphysik \& Cluster of Excelence PRISMA$^+$,\\ Johannes Gutenberg-Universitat Mainz, D-55099 Mainz, Germany
}
\affiliation[d]{School of Physics, Beihang University,\\ Beijing, 102206, China
}

\emailAdd{brenda@if.usp.br}
\emailAdd{amartine@if.usp.br}
\emailAdd{kanchan.khemchandani@unifesp.br}
\emailAdd{xiulei.ren@uni-mainz.de}
\emailAdd{lisheng.geng@buaa.edu.cn}

\abstract{In this presentation we discuss the generation of hadrons with an exotic quark content and hidden charm emerging from three-body interactions. To be more specific, we have studied the $KD\bar{D}^*/K\bar
D D^*$ and $ND\bar{D}^*/N\bar{D}D^*$ systems and predicted states generated from the respective three-body dynamics. In the case of the $KD\bar{D}^*/K\bar
D D^*$ system we predict the formation of a meson state with mass around $4307$ MeV and quantum numbers $I(J^P) = 1/2(1^-)$, $K^*(4307)$, and from the study of the $ND\bar{D}^*/N\bar{D}D^*$ system we found  $N^*$ states with masses in the range $4400\sim 4600$ MeV, width of $2\sim 20$ MeV and positive parity.\\
}

\FullConference{%
  XV International Workshop on Hadron Physics (XV Hadron Physics)
  13 -17 September 2021 \\
  Online, hosted by Instituto Tecnológico de Aeronáutica, São José dos Campos, Brazil\\
  }

\begin{document}
\maketitle

\section{Introduction}
In the past decades a large number of exotic states were observed experimentally due to the access available to higher energy regions at different facilities around the world, drawing a lot of attention to the subject. As a consequence there exist families of exotic states know as $X$, $Y$ and $Z$ (see, e.g., Refs. \cite{Hosaka:2016pey,Eberhard,Eulogio2016,stephen,richard}).

Exotic mesons, usually, can be understood as tetraquarks or as states originated from two-meson dynamics. Meanwhile, exotic baryons are usually described as pentaquarks or as states generated from baryon-meson dynamics. Most of the exotic states claimed recently have hidden or explicit heavy flavors (charm and bottom). 

In this talk we discuss the results obtained for two different three-body systems whose interaction generate exotic states: $KD\bar D^*/K\bar D D^*$ and $ND\bar D^*/N\bar D D^*$, both of them with hidden charm arising from the $D\bar{D}^*/\bar D{D}^*$ subsystem. The main motivation to investigate the $KD\bar{D}^*/K\bar{D}D^*$ system came from the fact that the subsystems $D\bar{D}^*$, $KD$, $KD^*$ have attractive interactions in s-wave from which the states $X(3872)$/$Z_c(3900)$, $D_{s0}(2317)$ and $D_{s1}(2460)$ arise, respectively \cite{Albaladejo:2018mhb,Bali:2017pdv,MartinezTorres:2014kpc,vanBeveren:2003kd,Guo:2006fu,Gamermann:2007fi,Barnes:2003dj,Gamermann:2006nm}. Also, if a state was to be found, the quantum numbers of the generated state (considering all interactions in s-wave) were expected to be compatible with a $K^*$ with a mass in the charmonium sector. Such a state would be a prediction that could encourage experimental searches for kaons at energies around $\sim 4-5$ GeV. Similarly, the study of the $ND\bar D^*$ system was inspired by the fact that the $ND$ and $ND^*$ subsystems are attractive and give rise, among others, to the state $\Lambda_c(2595)$ \cite{Nieves:2019nol,Liang:2017cji,Romanets:2012hm,Garcia-Recio:2008rjt,Mizutani:2006vq,Hofmann:2005sw}. Such attractive interactions, together with the attraction in the $D\bar{D}^*/\bar{D}D^*$ subsystem, lead us to contemplate that the formation of $N^*$ states with a three-body nature as a consequence of the $ND\bar D^*$ dynamics is quite probable.
Besides, the recent announcement of the LHCb collaboration claiming the existence of possible hidden charm pentaquarks with non-zero strangeness, masses around $4459$ MeV \cite{mwang} and quantum numbers yet to be determined further motivated us to carry out the calculations, since the masses found are close to the threshold of the $ND\bar{D}^*$ system. 
\section{Formalism}
The systems we are interested in are composed of three hadrons. To investigate their interactions we must obtain the corresponding $T$-matrix and this can be done by solving the Faddeev equation 
\begin{align}
    T &= T^1 + T^2 + T^3,\nonumber\\
    T^1 = t_1 + t_1G[T^2 + T^3],\phantom{na}\
    T^2 &= t_2 + t_2G[T^1 + T^3],\phantom{na}
    T^3 = t_3 + t_3G[T^1 + T^2],\label{faddeev}
\end{align}
with $G$ being a three-body loop function and $t_i$, $i=1,2,3$,  the two-body $t$-matrices describing the $(jk)$ subsystems [$j,k = 1,2,3$, $j\neq k \neq i$].

Solving the Faddeev equations, usually, is a cumbersome task. However, the three-body systems we are intending to study ($ND\bar{D}^*$ and $KD\bar{D}^*$) have an interesting characteristic that allows us to simplify the formalism for determining the $T$-matrix. When a three-body system,  composed by the particles $P_1$, $P_2$ and $P_3$, has $P_3$ lighter than the other two particles and $P_1 P_2$ cluster as a bound state/resonance, we can treat their interactions as those of a particle with fixed scattering  centers. As a consequence, Eq. (\ref{faddeev}) can be written as \cite{MartinezTorres:2020hus,Kamalov:2000iy,Deloff:1999gc,Chand:1962ec}
\begin{align}
  T = T_{31} + T_{32},\phantom{na}\text{with}\phantom{na} T_{31} = t_{31} + t_{31}G_3 T_{32},\phantom{na}&\phantom{na}T_{32} = t_{32} + t_{32}G_3 T_{31},\label{fad}
\end{align}
where $t_{31}$ ($t_{32}$) is the two-body $t$-matrix related to the subsystem $P_1P_3$ ($P_2P_3$) and $G_3$ the propagator of $P_3$ in the cluster.

Considering $K(N)$ as $P_3$, $D$ as $P_1$ and $\bar{D}^*$ as $P_2$ for the $KD\bar{D}^* (ND\bar{D}^*)$ system, with $D\bar{D}^*$ clustering as the states $X(3872)$ or $Z_c(3900)$ (with isospin $0$ or $1$, respectively), both our systems fulfil the the aforementioned criteria.

In this manner, to determine the three-body $T$-matrices for the $ND\bar{D}^*$ and $KD\bar{D}^*$ systems we must obtain $T_{31}$ and $T_{32}$, given in Eq. (\ref{fad}), for each system.
In the following we are going to succinctly present the formalism to do this. More details can be found in Refs. \cite{malabarba1,malabarba2}.

The elements $T_{31}$ and $T_{32}$ in Eq. ({\ref{fad}}) depend on the two-body $t$-matrices $t_{31}$ and $t_{32}$ as well as the loop function, $G_3$, of $P_3$ in the cluster. The latter propagator of $P_3$, with $P_3$ being $K$ ($N$) for the system $KD\bar{D}^*$($ND\bar
D^*$), is given by 
\begin{align}
    G_K &= \frac{1}{2M_a}\int \frac{d^3q}{(2\pi)^3}\frac{F_a(\mathbf{q})}{q_0^2 - \mathbf{q}^2 - m^2_K + i\epsilon},\phantom{na}
    G_N = \frac{1}{2M_a}\int \frac{d^3q}{(2\pi)^3} \frac{m_N}{\omega_N(\mathbf{q})}\frac{F_a(\mathbf{q})}{q_0 - \omega(\mathbf{q}) + i\epsilon}.\label{Gs}
\end{align}
The term $F_a$ in Eq. (\ref{Gs}) is a form factor related to the molecular nature of the cluster ($D\bar{D}^*$) and can be written as \cite{fator1,fator2,fator3} 
 \begin{eqnarray}
     F_a(\mathbf{q}) = \frac{1}{N}\int_{|\mathbf{p}|,|\mathbf{p - q}|<\Lambda} d^3\mathbf{p}\phantom{a}f_a(\mathbf{p})f_a(\mathbf{p -q}),\nonumber\\
     f_a(\mathbf{p}) = \frac{1}{\omega_{a1}(\mathbf{p})\omega_{a2}(\mathbf{p})}\cdot \frac{1}{M_a - \omega_{a1}(\mathbf{p}) - \omega_{a2}(\mathbf{p})},\label{FF}
 \end{eqnarray}
with $M_a$ being the mass of the cluster, $N = F_a(\mathbf{q} = 0)$ is a normalization constant, $\Lambda$ represents a cut-off $\sim$700 MeV and $\omega_{ai} = \sqrt{m_{ai}^2 + \mathbf{p}^2}$.

Next we need to calculate $t_{31}$ and $t_{32}$. The determination of $t_{31}$ and $t_{32}$ for the $KD\bar{D}^*$ and $ND\bar{D}^*$ systems is similar. Hence, to illustrate the procedure we consider, for example, the $KD\bar D^*$ system. First, let's determine the expression for $t_{31}$.

The $K(D\bar{D}^*)$ system, considering $D\bar{D}^*$ as $X(3872)$ or $Z_c(3900)$, has three possible isospin configurations: $\left|{KX,1/2,1/2}\right>$, $\left|KZ_c,1/2,1/2\right>$ and $\left|KZ_c,3/2,3/2\right>$ (where we use the notation $\left|A,I,I_3\right>$, with $A$ indicating the system, $I$ is the total isospin of the system and $I_3$ the corresponding isospin third component). Considering, for instance, the state $\left|{KZ_c,1/2,1/2}\right>$, the amplitude $t_{31}$ for $KZ_c\to KZ_c$ in isospin $1/2$ is given by 
\begin{align}
\left<{KZ_c,1/2,1/2}\right|t_{31}\left|{KZ_c,1/2,1/2}\right>.\label{KX1212}
\end{align}
Remembering that $t_{31}$ is related to the $KD$ subsystem, to determine the matrix element in Eq. (\ref{KX1212}) it is convenient to express the $\left|KZ_c\right>$ state in terms of the isospin of the $KD$ subsystem. To do this we use the Clebsch-Gordan coefficients
\begin{align}
    \left|{KZ_c,\frac{1}{2},\frac{1}{2}}\right> =\frac{1}{2}\left[\left|KD, 1, 1\right>\otimes \left|\bar{D}^*,\frac{1}{2},-\frac{1}{2}\right>+\frac{1}{\sqrt{2}}\left(\left|KD,1,0\right> + \left|KD,0, 0\right>\right)\otimes\left|\bar{D}^*,\frac{1}{2},\frac{1}{2}\right>\right].\label{clebesch}
\end{align}
Using Eq. (\ref{clebesch}) together with Eq. (\ref{KX1212}) we get that
\begin{equation}
    \left<{KZ_c,I=1/2,I_3=1/2}\right|t_{31}\left|{KZ_c,I=1/2,I_3=1/2}\right> \equiv {t_1}_{(22)} = \frac{1}{4}(t_{KD}^{I = 1} + 3t_{KD}^{I = 0}),\label{KXstate}
\end{equation}
where the $t_{KD}^{I = a}$ is the two-body $t$-matrix for the $KD$ subsystem with isospin $I = a$, $a = 0,1$, and the subscript $'(22)'$ in Eq. (\ref{KXstate}) stands for the transition $KZ_c\to KZ_c$, an element of the coupled channel $t$-matrix. Since the system $KX$ can also have total isospin $1/2$ it can couple to $KZ_c$, such that the transitions $KX\to KX$ ($t_{31(11)}$) and $KX\leftrightarrow KZ_c$ ($t_{31(12)}$ or $t_{31(21)}$) have to be considered in order to obtain $t_{31}$. The determination of the isospin weights for the transitions $KX\to KX$ and $KX\to KZ_c$ requires the exact same procedure as that for the case of $KZ_c\to KZ_c$, and the results are summarized in Table \ref{resultados1}.

\begin{table}[h!]
    \centering
    \begin{tabular}{c|c|c}
    &$KX$&$KZ_c$\\\hline
    $KX$&$\frac{1}{4}\left(3t_{KD}^{I=1} + t_{KD}^{I = 0}\right)$&$\frac{\sqrt{3}}{4}\left(t^{I = 1}_{KD} - t^{I = 0}_{KD}\right)$\\\hline
    $KZ_c$&$\frac{\sqrt{3}}{4}\left(t^{I = 1}_{KD} - t^{I = 0}_{KD}\right)$&$\frac{1}{4}\left(t_{KD}^{I = 1} + 3t_{KD}^{I = 0}\right)$ 
    \end{tabular}
    \caption{$t_{31}$ amplitudes for the $KD\bar{D}^*$ system in terms of the two-body $t$-matrices for the $KD$ subsystem.}
    \label{resultados1}
\end{table}
The computation of $t_{32}$ needs an analogous analysis to determine the contributions from interactions in different isospins and the results happen to coincide with those shown in Table \ref{resultados1}, except for a global minus sign to be considered for the non-diagonal terms and changing $D\to \bar{D}^*$.

Since $N$ has the same isospin as $K$ the results on the isospin weights obtained for $KD\bar{D}^*$ in Table \ref{resultados1} changing $K\to N$. The $t_{(32)}$ matrix for the $ND\bar{D}^*$ system is also analogous to the one obtained for the $KD\bar{D}^*$ system.

So, to determine the  $t_{31}$ and $t_{32}$ matrices we need to obtain the two-body $t$-matrices for the $KD$, $K\bar{D}$, $K\bar{D}^*$, $KD^*$, $ND$,  $N\bar{D}$, $N\bar{D}^*$ and $ND^*$ subsystems in different isospin configurations. 
 These amplitudes can be obtained by solving the Bethe-Salpeter equation
\begin{equation}
    t_{AB} = V_{AB} + V_{AC}G_{CC}t_{CB},\label{bethe}
\end{equation}
where $V_{AB}$ is the kernel for a channel made of two hadrons and can be obtained with an appropriate effective Lagrangian, and $G_{CC}$ is a two body loop function. The loop function $G$ is divergent and has to be regularized either with a cut-off or with dimensional regularization.

To obtain the two-body $t$-matrices $t_{DN}$ and $t_{\bar{D}^*N}$, we consider two distinct models: one based on the $SU(8)$ spin flavor symmetry \cite{simsu81} and other based on the $SU(4)$ and heavy-quark spin symmetries \cite{simsu41,simsu42}.

In the case of the $KD/K\bar{D}^*$ subsystems, to determine the $t_{KD}$ and $t_{K\bar D^*}$ we solve Eq. (\ref{bethe}) using as kernel the amplitude obtained from an effective Lagrangian based on heavy-quark spin symmetry~\cite{Burdman:1992gh,Weinberg:1991um,Wise:1992hn}.

Notice that the systems $ND$ and $ND^*$, unlike $KD$ $KD^*$, can be coupled in s-wave because of their quantum numbers [in s-wave the state $ND$ ($ND^*$) has spin-parity $J^P = 1/2^-$ ($J^P = 1/2^-, 3/2^-$)]. To consider the coupling between $ND$ and $ND^*$, in the $SU(4)$ model ~\cite{simsu42} a transition amplitude $DN\to D^*N$ is obtained through box diagrams (more details can be found in Ref. \cite{simsu42}).
In the $SU(8)$ model the $DN\to D^*N$ transition is described by a Weinberg-Tomozawa amplitude Ref.~\cite{simsu81}.


\section{Results}
\begin{figure}[h!]
    \centering
    \includegraphics[width=12cm]{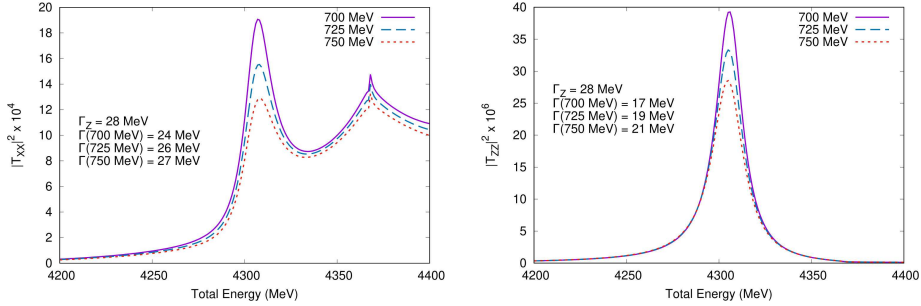}  
    \caption{Modulus squared of the three-body $T$-matrix for the transitions $KX\to KX$ (left) and $KZ_c\to KZ_c$ (right) considering the coupling between the $KX$ and $KZ_c$ channels.}
\label{resultadoscomloop}
\end{figure}

In figure \ref{resultadoscomloop} we show the results obtained for $|T|^2$ as a function of the energy of the  $KD\bar{D}^*$ system considering the configurations $KX$ and $KZ_c$. As we can see, in both processes, $KX\to KX$ and $KZ_c\to KZ_c$ [considering $KX$ and $KZ_c$ as coupled channels when solving Eq.~(\ref{fad})], a peak around $4300$ MeV shows up. The width of the $Z_c$ state was included in the formalism by implementing the transformation $M \to M - i\Gamma/2$, with $\Gamma\sim 28$ MeV, in the corresponding form factor of Eq.~(\ref{FF}).
When solving Eq.~(\ref{FF}) we have varied the cut-off  from $700$ to $750$ MeV, but no significant difference in the results were obtained from this variation (see Fig.  \ref{resultadoscomloop}).

It is interesting to note that the results for the $KX\to KX$ transition presented in Fig.~\ref{resultadoscomloop} show a second peak around $4375$ MeV which is related to the threshold of the three-body system.

When considering the $KZ_c\to KZ_c$ transition with isospin $3/2$ no signal of a bound state is found in $|T|^2$. Studies of further properties of $K^*(4307)$ can be found in Refs. \cite{Ren:2019umd,Ren:2019rts}. 

\begin{figure}[h!]
    \centering
    \includegraphics[width = 10cm]{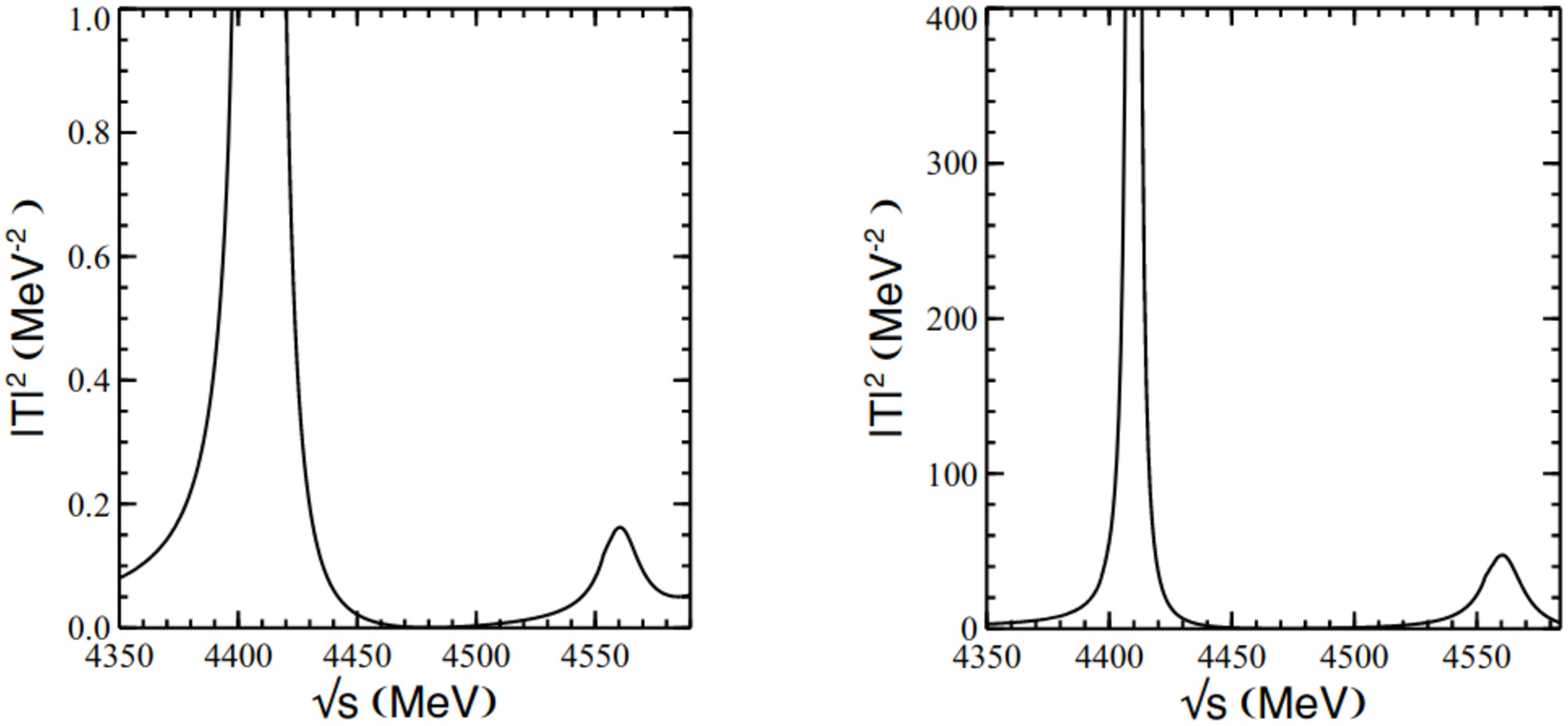}
    \caption{$|T|^2$ for the transitions $NX\to NX$ (left) and $NZ_c\to NZ_c$ (right) with $I(J^P) = 1/2\,(1/2^+)$ as functions of $\sqrt{s}$.}
    \label{resultadosN1212}
\end{figure}

\begin{figure}[h!]
    \centering
    \includegraphics[width =10cm]{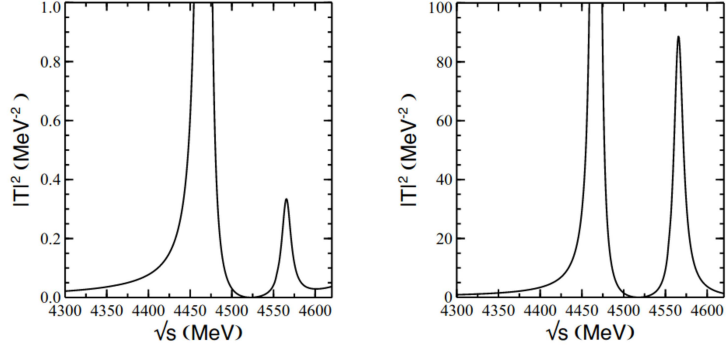}
    \caption{$|T|^2$ for the transitions $NX\to NX$ (left) and $NZ_c\to NZ_c$ (right) with $I(J^P) = 1/2(3/2^+)$ as functions of $\sqrt{s}$.}
    \label{resultadosN1232}
\end{figure}

Figure~\ref{resultadosN1212} shows the results for the  $NX\to NX$ and $NZ_c\to NZ_c$ transitions with $I(J^P) = 1/2\,(1/2^+)$ and Fig. \ref{resultadosN1232} shows the corresponding results in case of $I(J^P) = 1/2\,(3/2^+)$. In both cases we have considered the inputs of Refs. \cite{simsu41,simsu42} and $NX$ and $NZ_c$ as coupled channels.
As can be seem, the plots show that there are two peaks close to $4400$ MeV and $4550$ MeV for all four transitions.

The cut-offs used when calculationg Eq.~(\ref{FF}) vary from $700$ MeV to $770$ MeV and this variation causes a shift of $3-5$ MeV on the masses of the states obtained. 

The results for the $ND\bar{D}^*/N\bar{D}D^*$ system, obtained by considering different models to determine the two-body $t$-matrices of the subsystems ($SU(4)$ or $SU(8)$), are very similar and the small difference found from the differences in the two-body models, as well as the different cut-off used to solve Eq. (\ref{FF}), provide us uncertainties in the masses and widths of the states found. The results for the masses and widths of the $N^*$'s obtained together with the corresponding uncertainties are summarized in Table \ref{resultadossu4}.
\begin{table}[h!]
    \centering
    \begin{tabular}{c|c|c||c|c|c}
        Spin-parity & Mass(MeV) & Width (MeV)&Spin-parity & Mass(MeV) & Width (MeV) \\\hline
        $1/2^+$ & $4404 - 4410$ & $2$&$3/2^+$ & $4467 - 4513$&$\sim 3-6$\\
        $1/2^+$ & $4556 - 4560$&$\sim 4-20$&$3/2^+$ & $4558-4565$ & $\sim 5-14$
    \end{tabular}
    \caption{Masses and widths of the three-body $N^*$'s states found in the study of the $ND\bar D^*$ system.}
    \label{resultadossu4}
\end{table}

In case of $I=3/2$, the $ND\bar{D}^*/N\bar{D}D^*$ interaction is capable of generating $\Delta^*$ states with hidden charm and masses around $4359$ MeV with $\Gamma\sim 1.5$ MeV and $4512$ MeV with $\Gamma\sim 4$ MeV. For more details we refer the reader to Ref. \cite{Malabarba:2021taj}

\section{Conclusions and acknowledgments}
We conclude from this study that considering $D\bar{D}^*/\bar{D}D^*$ as a cluster and adding a nucleon or a kaon to it generates states with hidden charm and three-body molecular nature, i.e., states that are described in terms of hadrons interacting with each other while keeping their identities. 

The results obtained for the $KD\bar{D}^*/K\bar{D}D^*$ system show that a $K^*$ meson around $4307$ MeV should be observed in experimental investigation, 
while for the $ND\bar D^*/N\bar DD^*$ system $N^*$ states with $I(J^P)=1/2(1/2^+,3/2^+)$ and masses around $4400-4600$ MeV as well as $\Delta^*$'s with masses $4359-4512$ MeV and $I(J^P) = 3/2(1/2^+)$ are predicted.

This work is supported by the
 Conselho Nacional Cient\'{i}fico e Tecnol\'{o}gico (CNPq), grant number 305526/2019-7 and 303945/2019-2, and by the Funda\c{c}\~{a}o de Amparo \`{a} Pesquisa do Estado de S\~{a}o Paulo (FAPESP), processes numbers 2019/17149-3, 2019/16924-3 and 2020/00676-8.

\end{document}